\documentclass[final,mathptmx ]{aipproc}
\layoutstyle{8x11double}
\pdfoutput=1 

\newcommand{\BBless}{\ensuremath{\beta\beta0\nu}}

\newcommand{\elem}[2]{\ensuremath{{}^{#1}}#2}

\newcommand{\TO}{TeO$_2$~}

\newcommand{\tect}{\elem{130}{Te}~}

\begin{document}

\title{The low-temperature energy calibration system for the CUORE bolometer array}

% 07.20.Mc 	Cryogenics; refrigerators, low-temperature detectors, and other low-temperature equipment 
% 07.57.Kp 	Bolometers; infrared, submillimeter wave, microwave, and radiowave receivers and detectors           
% 14.60.Pq 	Neutrino mass and mixing                
% 87.56.bg      Radioactive sources 
              

%\author{<author3>}{
%  address={<common address for author2 and author3>}
%  ,altaddress={<author1 address>} % additional visiting address
%}

\author{S.~Sangiorgio}{
  address={University of Wisconsin, Madison, WI 53706, USA}
}

\author{L.~M.~Ejzak}{
  address={University of Wisconsin, Madison, WI 53706, USA}
}

\author{K.~M.~Heeger}{
  address={University of Wisconsin, Madison, WI 53706, USA}
}

%\author{J.~Houston}{
%  address={University of Wisconsin, Madison, WI, 53706 USA}
%}

%\author{K.~Kriesel}{
%  address={Physical Sciences Lab, Madison, Stoughton, WI 53589 USA}
%}

\author{R.~H.~Maruyama}{
  address={University of Wisconsin, Madison, WI 53706, USA}
}

\author{A.~Nucciotti}{
  address={Dipartimento di Fisica dell'Universit\`a di Milano-Bicocca, Milano I-20126, Italy}  
  ,altaddress={Sezione di Milano dell'INFN, Milano I-20126, Italy }
}

\author{M.~Olcese}{
  address={Sezione di Genova dell'INFN, Genova I-16146, Italy}
}

\author{T.~S.~Wise}{
  address={University of Wisconsin, Madison, WI 53706, USA}
}

\author{A.~L.~Woodcraft}{
  address={ SUPA, Institute for Astronomy, Edinburgh University, Blackford Hill, Edinburgh EH9 3HJ, UK}
}

%\author{D. Zou}{
%  address={University of Wisconsin, Madison, WI, 53706 USA}
%}

\begin{abstract}
The CUORE experiment will search for neutrinoless double beta decay (\BBless) of \tect using an array of 988 \TO bolometers operated at 10 mK in the Laboratori Nazionali del Gran Sasso (Italy). The detector is housed in a large cryogen-free cryostat cooled by pulse tubes and a high-power dilution refrigerator. 
The \TO bolometers measure the event energies, and a precise and reliable energy calibration is critical for the successful identification of candidate \BBless~and background events.
The detector calibration system under development is based on the insertion of 12 $\gamma$-sources that are able to move under their own weight through a set of guide tubes that route them from deployment boxes on the 300K flange down into position in the detector region inside the cryostat. 
The CUORE experiment poses stringent requirements on the maximum heat load on the cryostat, material radiopurity, contamination risk and the ability to fully retract the sources during normal data taking. Together with the integration into a unique cryostat, this requires careful design and unconventional solutions. 
We present the design, challenges, and expected performance of this low-temperature energy calibration system.
\end{abstract}

\maketitle

%%%%%%%%%%%%%%%%%%%%%%%%%%%%%%%%%%%%%%%%%%%%
%% MAINMATTER
%%%%%%%%%%%%%%%%%%%%%%%%%%%%%%%%%%%%%%%%%%%%
%\linenumbers

\section{Introduction}
%%%%%%%%%%%%%%%%%%%%%%%

% CUORE, the need of a calibration system, requirements
In the CUORE experiment \cite{CUORE-NIMA}, an array of 988 \TO bolometers will be used to search for neutrinoless double beta decay (\BBless) with the aim of investigating the nature and the absolute mass of neutrinos. CUORE bolometers are operated as thermal calorimeters at a temperature of $\sim$10~mK. The energy released by a particle is converted into phonons and the subsequent temperature rise is read out by a thermistor and converted into a voltage pulse, whose amplitude is related to the particle energy in a nonlinear way that must be determined experimentally for each bolometer. An energy calibration is necessary to reliably establish the bolometer response over the entire energy spectrum, in particular close to the region of interest for  \BBless~of \tect around 2527~keV. 
% In the last steps of the analysis of the data of our bolometers, for each bolometer, the energy spectra of each dataset are summed together. A second sum over all detectors is then performed. Results for \BBless are obtained from the study of the energy window of ~100 keV around 2530 keV in this final spectrum.
 Any uncertainty in the calibration will affect the energy resolution of the bolometers (thus worsening our sensitivity) and introduce a systematic error in the search for \BBless~events.

The bolometers will be cooled down in a large cryogen-free cryostat \cite{Nucciotti-LTD12} where the 4K stage is achieved by means of 4 pulse tubes and the base temperature by a high-power dilution refrigerator. The design of such a cryostat is challenging, mainly because of the size and weight of the detector and the need to include $\sim$15~t of lead for radioactive shielding inside the cryostat. 
% As we will see later, the calibration system will be tightly integrated with the cryostat and will therefore share some of the challenges.

Several different requirements must be satisfied in the design of the detector calibration system. We must 
\begin{itemize}
\item not exceed the maximum event rate of 150~mHz on each bolometer, to avoid pile-up and baseline build-up due to the intrinsic slow response of bolometers;
\item meet the maximum allowed heat load at each stage of the cryostat;
\item have the ability to change the sources and the calibration source isotope;
\item prevent the calibration time from significantly affect detector live time;
\item use only certified materials with low radioactivity, store calibration sources outside the cryostat during normal data-taking, and otherwise avoid any risk of radioactive contamination, since \BBless~observation requires an ultra-low-background environment;
% may need to quantify the radioactivity limits?
\item operate safely and reliably for the entire experiment lifetime (> 5 years).
\end{itemize}

%The design of the detector calibration system must satisfy several different requirements:
%\begin{itemize}
%\item must not exceed the maximum event rate of 150~mHz on each bolometer, to avoid pile-up and baseline build-up due to the intrinsic slow response of bolometers
%\item meet the maximum allowed heat load at each stage of the cryostat
%\item ability to change the sources and the calibration source isotope
%\item calibration time does not significantly affect detector live time
%\item use only certified materials with low radioactivity, store calibration sources outside the cryostat during normal data-taking, and otherwise avoid any risk of radioactive contamination, since \BBless~observation requires an ultra-low-background environment
%% may need to quantify the radioactivity limits?
%\item safe and reliable operation for the entire experiment lifetime (> 5 years)
%\end{itemize}

This paper describes the calibration system that is being developed to calibrate the CUORE detector according to these requirements.

\section{Design of the calibration system}
%%%%%%%%%%%%%%%%%%%%%%%

%\begin{figure}
% \includegraphics[width=\columnwidth]{fig/DCS-onepage-overview}
%\caption{Some images illustrating the concept of the CUORE calibration system. Left: 3D model of the cryostat flanges with the calibration motion boxes on top and the 12 guide routings integrated. Left top: 3D rendering of a drive spool assembly (view from inside of motion box). Center: Sketch of the source carrier and photo of a prototype. Bottom center: From the GEANT4 simulations, top view of calibration source layout with respect to the crystal towers (sources enlarged for visibility). Right: Schematic view of the guide tubes' material and thermal couplings with the cryostat flanges.}
% \label{fig:overview}
%\end{figure}

\begin{figure}
 \includegraphics[width=0.4\columnwidth]{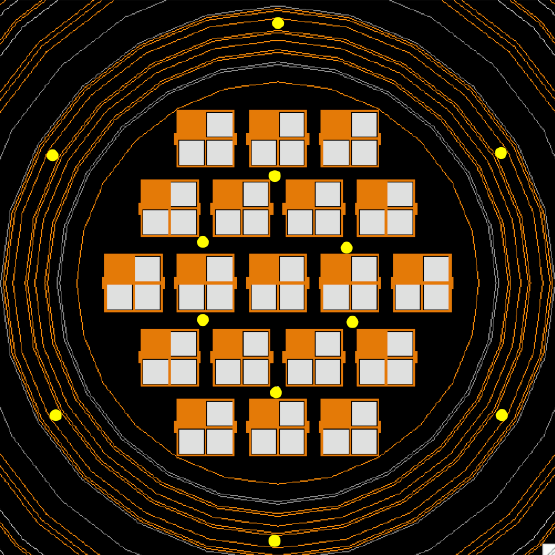}
\caption{Top view of calibration source layout with respect to the crystal towers (sources enlarged for visibility). Visualization from GEANT4.}
 \label{fig:source-locations}
\end{figure}

The energy calibration of the CUORE bolometer array will be based upon regular measurements with $\gamma$-sources of well-known energies. This is similar to what was done in the Cuoricino experiment \cite{QINO-PRC78}, the predecessor to CUORE. For best results, five or more lines should be clearly visible in each bolometer spectrum. The CUORE detector consists of a tightly-packed array of 19 towers with 13 planes of four bolometers each. To overcome the detector self-shielding, calibration sources need to be placed between the towers. Fig.  \ref{fig:source-locations}  shows the final arrangement of the 6 internal and 6 external sources. They were optimized with GEANT4 Monte Carlo simulations to produce a uniform illumination of all detectors. 

The CUORE calibration system is based upon 12 radioactive source strings that are able to move, under their own weight, through a set of guide tubes that route them from outside the cryostat to their locations around and between the bolometers. The system consists also of 4 computer controlled deployment boxes above the 300K flange, cooling mechanisms to thermalize the source strings at 4K, and a vacuum and purge system.

\begin{figure}
 \includegraphics[width=0.60\columnwidth]{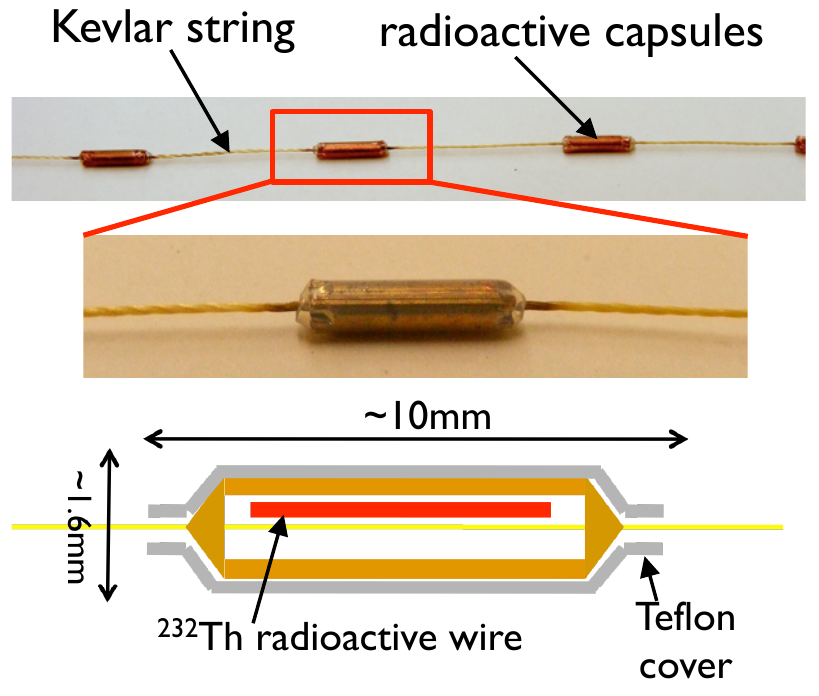}
\caption{Schematic of the source carrier and photo of a prototype source string.}
 \label{fig:source-string}
\end{figure}

As shown in Fig. \ref{fig:source-string}, the source string is made of 30 copper crimp capsules, evenly spaced over a length equal to the detector height (85 cm), at the bottom of a Kevlar string (0.35~mm diameter). Each capsule is 8~mm long, has a 1.6~mm outer diameter (OD) and houses a radioactive thoriated Tungsten wire. A heat-shrunk PTFE cover is placed over each capsule to minimize the friction against the guide tube in which the string is moving. The activity of each capsule will be 130~mBq for the internal sources and 633~mBq for the external ones. 
% THERE IS NO SPACE TO EXPLAIN ABOUT THE SOLID ANGLE PROBLEM

\begin{figure}
 \includegraphics[width=\columnwidth]{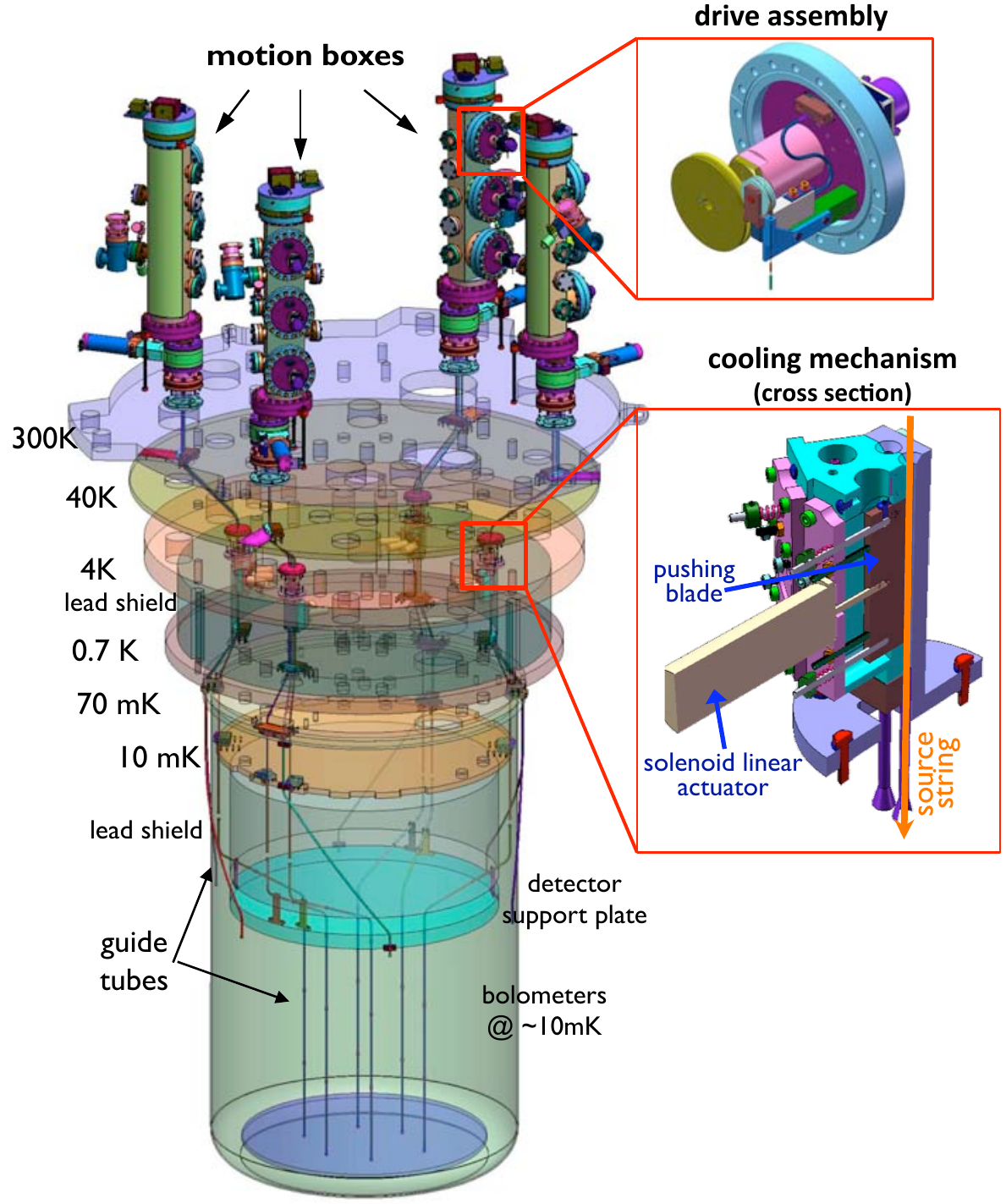}
\caption{3D model of the cryostat flanges with the calibration motion boxes on top and the 12 guide tube routings integrated. On the right, 3D rendering of a drive spool assembly (view from inside of motion box) and of the cooling mechanism.}
 \label{fig:3D}
\end{figure}

A storage and deployment mechanism for the source carriers will be provided by four motion boxes that sit on top of the cryostat, each of which will host three independently-controlled drive spool assemblies (Fig. \ref{fig:3D}).  The insertion and extraction of the calibration sources will be controlled remotely  by a computer control system which will be integrated into both the cryostat's slow control and the experiment's database.

The source strings are routed by guide tubes from above the 300K flange through the cryostat's flanges (Fig. \ref{fig:3D}). The guide tubes also provide a thermal connection to the various stages of the cryostat. The presence of the lead shields and other cryostat subsystems forces the tubes to bend in several places.  In addition, manufacturing and cleaning constraints force us to split them into several sections.

Above the mixing chamber, where there is no direct line of sight with the detector because of the presence of the a lead shield, stainless steel (type 304) is used for the tubing (ID 5~mm, OD 5.4~mm).
In the lower parts, tubes (ID 4~mm, OD 2~mm) will be machined out of bars of a special high-purity, oxygen-free copper.  
Thermal and mechanical connections are established by copper mounts to all cryostat flanges. A set of thermometers  will be placed at various points along the guide tubes to monitor the temperature during the source insertion and extraction.

\section{Thermal analysis}
%%%%%%%%%%%%%%%%%%%%%%%

In the context of the cryostat design \cite{Nucciotti-LTD12}, a thermal analysis of the cryostat was performed and the thermal budget at each stage calculated and distributed among the various subsystems. The cooling power available at each thermal stage for the calibration system evaluated under static equilibrium conditions is shown in Table \ref{tab:table-thermal-calc}.  

\begin{table}[tbp]
\begin{tabular}{cccc}
\hline 
Stage & calibration cooling & heat load from  & radiated power from \\
          & power budget &  guide tube conductivity & source strings at 4~K \\
\hline 
40 K & $\sim 1$ W & $\sim 1$ W & --  \\
4 K & 0.3 W & 0.02 W & --  \\
0.7 K & 0.55 mW & 0.13 mW & 0.08 $\mu$W \\
70 mK & 1.1 $\mu$W & negligible & 0.3 $\mu$W   \\
10 mK & 1.2 $\mu$W & 1.07 $\mu$W & 0.08  $\mu$W  \\
detector & $<$ 1 $\mu$W & --  & 0.25 $\mu$W \\
\hline
\end{tabular}
\caption{Cooling power available to the calibration system at each thermal stage of the cryostat and heat load contributions from the calibration system parts and operations.}
\label{tab:table-thermal-calc}
\end{table}

Studies to understand the dynamic behavior of the cryostat in response to thermal excitation have also been done but, given its complexity, the real behavior is largely unknown. As a conservative approach the calibration system is designed so that the thermal load at any moment is lower than the maximum allowed in static conditions. This will ensure that the working points of the bolometers are not affected by the calibration.

A complete thermal analysis of the calibration system was performed, trying to include all relevant sources of heat load. We will briefly discuss them here. The calculated heat loads are summarized in Table \ref{tab:table-thermal-calc}.

\begin{figure}
 \includegraphics[width=.74\columnwidth]{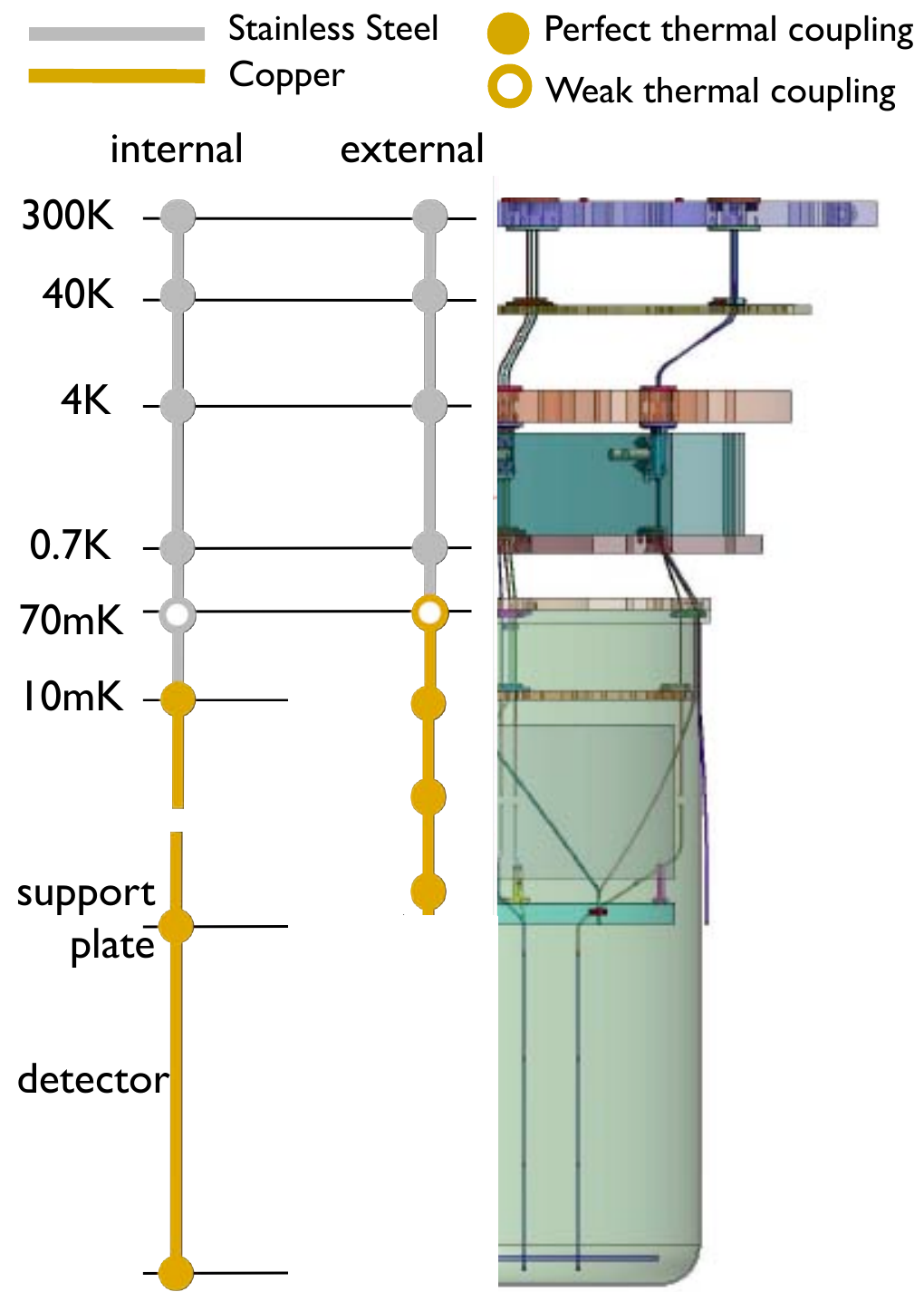}
\caption{Schematic view of the guide tubes' material and thermal couplings to the cryostat flanges.}
 \label{fig:guide-tubes}
\end{figure}

{\bf Thermal conductivity of guide tubes} \\ 
A schematic model of the thermal connections is shown in Fig.~\ref{fig:guide-tubes}. By using data from literature \cite{lounasmaa,nist} we calculated the heat load from the guide tubes' conductivity.
As shown in Table \ref{tab:table-thermal-calc}, all values are within the allowed limit if good thermal contact is made at all stages, with the exception of the 70~mK plate where a weak coupling is needed because of the small distance (11~cm) from the 0.7 K plate. Weak thermal coupling will be achieved by introducing low--conductivity spacers in the tube mounts.

{\bf Radiation heat inflow down the guide tubes}\\
The radiation from the motion box at room temperature will inevitably be funneled down the guide tubes and contribute to the heat load on the various stages. The bends along the path are thought to have a similar effect as typical shields or baffles. Worst-case calculations have shown that this heat contribution is expected to be negligible. If this proves not to be the case, we can mitigate unwanted radiation by blackening the inner surface of the section of the guide tube between 300~K and 4~K. 

{\bf Heat radiated by the source strings} \\
In order to meet the maximum heat load in the detector region, calculations show that the source strings cannot have a temperature higher than 4~K when they are fully deployed next to the bolometers (i.e. during calibration measurements). It is worth noting that the guide tubes will shield the radiant heat from the detectors, which would otherwise be immediately warmed up and become unusable.  The source carrier design has the advantage of being small and light, thus reducing the total heat to be removed to cool it down to 4~K. Nevertheless, since  radiation cooling is ineffective, a mechanical system that provides good thermal contact between the capsules and a heat sink at 4~K is currently being developed. The concept is based upon a linear solenoid actuator operated at 4~K that squeezes the capsules against a heat sink at 4~K. A conceptual design drawing is shown in Fig. \ref{fig:3D}. Although preliminary tests at room and liquid nitrogen temperature are encouraging, cooling the sources is still one of the most critical open design issues.  

{\bf Thermal conductivity of the source strings}\\
The source strings are connected to 300~K at all times in the motion boxes. Due to their size and the relatively small thermal conductivity of Kevlar \cite{nist}, the heat load will be negligible if we are able to thermalize the string at each thermal stage. Unfortunately, the thermal contact between the string and the guide tube is ill-defined. In the worst case scenario, when a source is fully deployed, the heat from the string conductance will overcome the heat loss by radiation. Thus it will effectively warm up the source string above its nominal operating temperature of 4~K. Therefore, the source carrier strings need to be effectively thermalized at least at the 4K stage.

{\bf Friction heat during source string motion}\\
A feature of the calibration system design is that there are 12 strings that move over 2.5 meters from 300~K into a region at 10 mK. During insertion and extraction, the source strings  produce friction by sliding against the tubes. This is especially critical at the bends where the friction force depends exponentially on  the bending angle and the friction coefficient. 

We have developed a simulation that models the friction and adjusts the source extraction speed so that the heat dissipated by friction on each stage does not overcome the maximum allowed in static conditions. Source string motion as slow as 0.1 mm/s is needed to meet this constraint. The time required for the extraction of the sources is then evaluated based on pairing one internal string with an external one and then staggering the pairs, resulting in $\sim$48 hours extraction time. During the commissioning tests that will be performed in the CUORE cryostat, we plan to study the dynamic response of the cryostat to excess heat loads and evaluate the possibility of moving the source faster depending on the recovery time constant of cryostat and detector.

To reduce friction, we are evaluating the use of PTFE for the bent sections of the guide tubes. We are also investigating materials other than Kevlar for the source strings, with a lower friction coefficient and similar mechanical properties.

%{\bf Heat load due to residual He gas in the MC}\\
%Although its real pressure is largely unknown, residual gas He in the MC could introduce a significant conductivity 
%between the source string (T$\sim$4~K or lower) and the detectors. However, geometrical considerations (i.e. source surrounded by guide tubes) and calculations showed that this type of heat load will be negligible.

\section{Prototyping and testing}
% results from lab measurements
For room temperature motion tests, a representative full-length guide tube routing was assembled together with a drive assembly. The source is moved by a stepper motor while its tension is continuously monitored by a miniature load cell.  An optical encoder, a proximity sensor and a camera are also present to monitor the source position. Dummy sources were manufactured for the tests. A LabView program has been developed to control the motor and acquire data from all sensors.

Our testing so far has demonstrated that, at room temperature, the source moves along the guide tubes in a reliable and reproducible way. The load cell reading shows a pattern that reflects the source string's path through the guide tubes and can be used to monitor the source for problems. Positioning accuracy of about 0.5 mm can be achieved. A wear test was performed, which showed that only small fraying occurs in the Kevlar string after more than 10000 spooling cycles.

\section{Conclusions}
The design of a calibration system for CUORE that allows radioactive sources to be inserted into and extracted from a cryostat is being finalized. The cryostat for CUORE will be installed at the beginning of 2010, along with a full calibration routing for testing purposes. Final installation and commissioning of the calibration system will take place in the second half of 2010. 

%%%%%%%%%%%%%%%%%%%%%%%%%%%%%%%%%%%%%%%%%%%%%%%%
%% BACKMATTER
%%%%%%%%%%%%%%%%%%%%%%%%%%%%%%%%%%%%%%%%%%%%%%%%

\begin{theacknowledgments}
We gratefully acknowledge the support of the University of Wisconsin and the U.S. Department of Energy, Office of Science, Nuclear Physics through OJI Grant DR-FG02-08ER41551. We also thank Ken Kriesel, Jackie Houston and Dan Zou  for their contributions to the design and testing of this system. This work was done in collaboration with and as part of the CUORE experiment.
\end{theacknowledgments}

\bibliographystyle{aipproc}   % if natbib is available

\end{document}